\documentclass[prb,showpacs]{revtex4} 
\usepackage{graphicx}                 
\begin{document}

\title{Thermal conduction of one-dimensional isotopically disordered harmonic lattice.}

\author{Vladimir~N.~Likhachev}%
\email{vl@deom.chph.ras.ru} %
\affiliation{Institute of Biochemical Physics, Russian Academy of Sciences,
ul.Kosygina 4, Moscow 119991 GSP-1, Russia. }

\author{Juraj~Szavits-Nossan}%
\email{jszavits@inet.hr}%
\affiliation{Department of Physics, Faculty of Science, University of Zagreb,
\\ P.P.~331, HR-10002 Zagreb, Croatia.}

\author{George~A.~Vinogradov}%
\email{gvin@deom.chph.ras.ru}%
\affiliation{Institute of Biochemical Physics, Russian Academy of Sciences, ul.
Kosygina 4, Moscow 119991 GSP-1, Russia, \\ E-mail: gvin@deom.chph.ras.ru}


\baselineskip = 18 pt

\begin{abstract}

In the present communication we consider the one-dimensional (1D) isotopically
disordered lattice with the harmonic potential. Our analytical method is
adequate for any 1D lattice where potential energy can be presented as the
quadratic form $U=\frac12 \sum_{i,j} q(i) \, U_{ij} \, q(j)$, where $q(i)$ --
coordinate or velocity of $i$-th particle. There are derived the closed system
of equations for the temporal behavior of the correlation functions. The final
expressions allow to calculate the kinetics and dynamics of the system --
energy, temperature profile, thermal conduction and others. There is developed
the method for the calculation of the evolution of the eigenvalues
(frequencies) and eigenvectors (relaxation times) to their stationary values.
Exact results are obtained for times $\simeq 10^{14}$. The methods are
suggested allowing to extend the range of the relaxation times  upto $\simeq
10^{28}$. The spectrum of relaxation times reaches it constant value starting
from the number of particles $N$ in the lattice $N \geq 300$. Thermal
conductance $\kappa$ has the non-monotonic character: for the number of
particles $N < 300$ $ \kappa$ increases as $\kappa \simeq 2.4 \, \lg \, N$,
reaches the maximum value equal $\simeq 4.0$ at $N \simeq 300$ and then slowly
decreases upto $N = 700$.  The stationary state is unique and satisfies the
Gibbs distribution. An excellent agreement between numerical simulations and
analytical results is obtained where possible.

\end{abstract}


\maketitle


At present, more attention is focused on statistical and transport properties
of low-dimensional systems \cite{Lep03a}. Besides obvious achievements there
exists few unresolved fundamental problems, e.g. if there exists the unique
stationary state? If this state exists, then what is the time of relaxation to
stationary state ? Are some values self-averaged,  i.e. if the thermodynamical
limit really exists for these values? What are the sufficient conditions for
the observation of finite value of thermal conductance in low-dimensional
systems?

The diverging value of thermal conductance $\kappa \propto N^{\alpha}$ with
$0.17 \leq \alpha \leq 0.5$ was obtained for certain model systems \cite{Lep98,
Lep97, Sav02, Hat99, Dha01a}. Theoretical estimations for $\alpha$ give values
2/5 \cite{Lep98, Ern91} or 1/3 \cite{Nar02} depending on the chosen model.

There was obtained the final value of thermal conductivity \cite{Sav02, Gar01,
Gia00, Gen00}, and, moreover, numerical simulations predict even the ``phase
transition`` from normal to diverging thermal conductivity when  temperature
and/or parameters vary. However some of these results were criticized
\cite{Yan03}.

Usually it is supposed that non-integrability is the necessary but not enough
condition for the final value of thermal conduction in 1D systems
\cite{Dha01b}. There was demonstrated \cite{Lep98} that any 1D system with the
acoustical branch of excitation should have an infinite thermal conductivity in
the limit of low temperatures. This statement is supported by the theorem that
the one-dimensional system with momentum conservation and nonzero pressure have
infinite thermal conductivity \cite{Pro00}.

Many model lattices with different potentials of interaction were
investigated. Important qualitative and quantitative results were
obtained, though controversial results were observed for the same model
systems. These and associated problems are thoroughly considered in the
recent Review \cite{Lep03a}.

1D lattices are very useful prototypes for the (analytical and numerical)
investigation of kinetic, dynamical and transport properties of more complex
and practically interesting systems such as carbon nanotubes (see e.g.
\cite{Mar00, Mar02, Mar03, Cao04}).

The problem of the dynamics of the disordered lattices was formulated over 50
years ago by F.Dyson \cite{Dys53}, who gave the general formulation of the
problem.

Dyson's problem was chosen for the thorough investigation as this system meet
most of the problems more or less common in the simulation of thermal
conductance in low-dimensional systems. Small number of parameters allows to
analyze their influence on the final results, and dynamical and transport
properties.


We start from the quadratic hamiltonian with fixed boundary conditions (we are
mainly interested in methodology, and the role of boundary conditions and other
parameters will be considered and presented in a separate publication.)
\begin{equation}
\label{our_ham}%
H =\displaystyle\frac{1}{2}%
\sum_{i=1}^N \, m(i) \, v^2(i) \, - \displaystyle\frac{1}{2} \,
\sum_{i=1}^N \,  x(i) \, U_{ij} \, x(j) , %
\end{equation}
where in the considered case the matrix ${\hat U}$ has the form:
\begin{equation}
\label{Matrix_U}%
{\hat U} =
\left(%
\begin{array}{cccccc}

-2  & 1  & 0      & \cdots & 0      & 0 \\
  1 & -2 & 1      & \cdots & 0      & 0 \\
  0 & 1  & -2     & \cdots & 0      & 0 \\
  . & .  & .      & .      & .      & . \\
  0 & 0  & \cdots & 1      & -2     & 1 \\
  0 & 0  & \cdots & 0      & 1      & -2 \\
\end{array}%
\right).
\end{equation}
Recall that our approach is valid for the arbitrary quadratic Hamiltonians of
the form
\begin{equation}
\label{Quad_ham}%
H = \sum \, q(i) \, H_{ij} \, q(j),
\end{equation}
giving rise to the linear equations of motion, where $q(i),\, q(j)$ --
coordinates or velocities of the particles.

For the modelling of the heat reservoir we use the Langevin random forces with
friction, where in addition to dynamical equations the members [$ -\gamma(i) \,
\dot x(i) + \xi(i) $] were added, and $\gamma(i), \,\, \xi(i)$ -- the
"friction" coefficient and random force, correspondingly. Lets consider first
the most general case, when the Langevin sources acts on all particles of the
lattice. Then the equation of motion in matrix form is
\begin{equation}
\label{Eq_mot_1}%
{\dot{\vec q}} =  \left( {\hat A} \, {\vec q} + {\vec \Xi} \right),
\end{equation}
where time dependent vector of the state ${\vec q}$ is represented for
convenience in the form
\begin{equation}
\label{vec_q}%
{\vec q}(t) =
\left(%
\begin{array}{c}
  {\vec V} \\
  {\vec X} \\
\end{array}%
\right)%
 = \{ v(1), \, v(2), \, \ldots \, v(N); x(1), \, x(2), \, \ldots, x(N) \},
\end{equation}
vector ${\vec\Xi}(i) = \{ \xi(1), \, \xi(2), \, \ldots, \xi(N) \}$   in
(\ref{Eq_mot_1}) is the vector of uncorrelated random forces, where random
forces $\xi(i)$ obey the standard relations:
\begin{equation}
\label{xi_xi}%
\left\langle \xi(i, \,t_1) \, \xi(j, \,t_2) \right\rangle = 2 \, \delta_{ij} \,
\delta(t_1 - t_2) \, \gamma(i) \, T(i)/{m^2(i)}
\end{equation}
Matrix ${\hat A}$ has the block form
\begin{equation}
\label{Matrix_A}%
{\hat A} =
\left(%
\begin{array}{cc}
  {{\hat M}^{-1} \, {\hat U}} & {-{\hat M}^{-1} \, {\hat \Gamma}} \\
  {\hat I}                    & {\hat 0} \\
\end{array}%
\right),
\end{equation}
and matrix ${\hat U}$ namely this one that used in (\ref{our_ham}) and
(\ref{Matrix_U}).

In equation (\ref{Matrix_A}) ${\hat M}(i,j) = \delta_{ij} \, m(i)$ is the
diagonal matrix of random masses of particles, and ${\hat M}^{-1}$ -- inverse
matrix. We have chosen well known from literature the mass distribution such
that randomly and with the equal probability masses have values  1 or 1/2.
Matrix ${\hat \Gamma} = \delta_{ij} \, \gamma(i)$ -- diagonal matrix of
friction coefficients in Langevin forces for the corresponding particles,
matrices ${\hat I}$ and ${\hat 0} $, are, correspondingly, unit and null
matrices.

Generally speaking, matrix ${\hat A}$ falls into a category of random matrices
widely used in mathematics and physics (see, e.g. recent review \cite{For03}).

Initial condition is chosen in the form ${\vec q}(t=0) = {\vec 0}$, i.e.
initial temperature of all particles are equal to zero. Then the formal
solution of equation (\ref{Eq_mot_1}) has the form
\begin{equation}
\label{Eq_q}%
{\vec q}(t) = \int_0^t \, {\hat W}(t-\tau) \, {\hat \Xi}(\tau) \, d\tau
\end{equation}
Note, that if the initial conditions are arbitrary, i.e. ${\vec q}(t=0) = {\vec
q}_0$, then there  the solution of the inhomogeneous solution should be added
to the r.h.s. of (\ref{Eq_q}) and it has a form:
\begin{equation}
\label{Eq_W_1}%
{\vec q}(t) = {\hat W}(t) \, {\vec q}_0 + \int_0^t \, {\hat W}(t-\tau) \, {\hat
\Xi}(\tau) \, d\tau.
\end{equation}
The time evolution operator ${\hat W}(t)$ satisfies the equation
\begin{equation}
\label{Eq_w}%
\displaystyle\frac{d \, {\hat W}}{d \, t} = {\hat A} \, {\hat W}, \qquad {\hat
W}(t=0) = {\hat I}
\end{equation}
Define now the $2N \times 2N$ matrix of correlation functions (keeping in mind
the definition (\ref{Eq_q}) )
\begin{equation}
\label{Corr_func_1}%
\left\langle q(i,\,t) \, q(j,\,t) \right\rangle = \int_0^t \int_0^t d\tau_1 \,
d\tau_2 W_{i,i_1}(t-\tau_1) W_{j,j_1}(t-\tau_2) \, \left\langle
\xi_{i_1}(\tau_1) \, \xi_{j_1}(\tau_2) \right\rangle.
\end{equation}
Using the expression (\ref{xi_xi}), one can get:
\begin{equation}
\label{Corr_func_3}%
C_{ij} \equiv \left\langle q(i, \,t) \, q(j, \,t) \right\rangle = \sum_p
\displaystyle\frac{2 \, \gamma(p) \, T(p)}{m^2(p)} \, \int_0^t \, d\tau \,
W_{ip}(\tau) \, W_{jp}(\tau).
\end{equation}
where $p$ is the number of a particle in the lattice subjected to Langevin
forces.

Expression for $W_{ip}(\tau)$ describes the evolution of the hole system, when
at $t=0$ the velocity of $p$-th particle is equal to unity, and displacement
from the equilibrium position is zero. All other particles have zero
displacements and velocities.

In the general case the evolution operator $W_{ip}(t)$ for the $p$-th particle
forecasts its temporary behavior at the initial condition $W_{ip}(t=0) =
\delta_{ip}$:
\begin{equation}
\label{W1}%
\displaystyle\frac{d \, W_{ip}}{dt} = \sum_j \, A_{ij} \, W_{jp}.
\end{equation}
Generally speaking, one can solve the system of $2 \, N$ equations (\ref{W1})
for every of Langevin particles $p$. As it can be seen from
(\ref{Corr_func_3}), the contributions from Langevin sources are also
independent and additive. It also means the linearity of results on
temperatures of Langevin sources.

The standard approach for the solving the system (\ref{W1}) consists in
transformation to the problem for eigenvalues. Namely this approach was used in
this communication. The solution will be searched as the series by
eigenfunctions of operator  $\hat A$.

Let's consider the problem for the eigenvalues of relaxing vibrational states
(vibrations relax because of an existence of friction in Langevin forces):
\begin{equation}
\label{Vib_1}%
{\hat A} \, {\vec q}_k = \lambda_k \, {\vec q}_k,
\end{equation}
where $k$ is the number of eigen solution (vibration). The complex value
$\lambda_k$ can be represented in the form:
\begin{equation}
\label{Lamb}%
\lambda_k = -\displaystyle\frac{1}{\tau_k} + i \, \omega_k,
\end{equation}
where ${\tau_k}$ and $\omega_k$ are, correspondingly, the relaxation time and
eigen frequency of $k$-th mode. $\lambda_k$ is not purely imaginary because of
the relaxation in Langevin forces, and formally because the matrix ${\hat A}$
is the unsymmetrical matrix and its eigenvalues in the general case are complex
values.

Let's expand $W_{ip}$ in the series over eigenvalues $q_k$:
\begin{equation}
\label{Ser_1}%
W_{ip} = \sum_{k=1}^{2\,N} \, e^{\lambda_k \, t} \, c_{kp} \, q(i)_k
\end{equation}
Actually we have $N$ equations for all Langevin particles with still unknown
coefficients $c_{kp}$, which can be determined from the same equation
(\ref{Ser_1}) at $t=0$ and the initial condition $W_{ip}(t=0) = \delta_{ip} $:
\begin{equation}
\label{Coeff_1}%
\delta_{ip} = \sum_{i=1}^{2 \, N} \, c_{kp}  \, q(i)_k,
\end{equation}
what in the symbolic form can be rewritten as
\begin{equation}
\label{second}%
{\hat B} = {\hat A} \, {\hat x}, \quad \mbox{где} \qquad {\hat B} =
\delta_{ip}, \,\, {\hat x} = c_{kp}, \,\, {\hat A} = q_k(i).
\end{equation}

Substituting the solution of the equation (\ref{Ser_1}) in (\ref{Corr_func_3}),
one finally gets
\begin{equation}
\label{Fin_1}%
C_{ij} = \sum_{p=1}^N \displaystyle\frac{2 \, \gamma(p) \, T(p) }{m^2(p)}
\sum_{k_1, \, k_2 = 1}^{2\,N} \, \displaystyle\frac{1 - e^{(\lambda_{k_1} +
\lambda_{k_2}) \, t}}{\lambda_{k_1} + \lambda_{k_2}} \, q(i)_{k_1} \,
q(j)_{k_2}
\end{equation}

At  $t \to \infty$ the fraction in (\ref{Fin_1}) reduces to $1/(\lambda_{k_1} +
\lambda_{k_2})$

One can check, that our results (\ref{Corr_func_3}) and (\ref{Fin_1}) satisfies
the equation for the correlation function, which can be obtained from the
Fokker-Planck equation for the distribution function $P({\vec q})$ with the
initial condition $P({\vec q}, t=0) = \delta({\vec q})$.

An expression (\ref{Corr_func_3}) is easily generalized for the correlation
functions taken at different instants of time:
\begin{equation}
\label{Corr_func_4}%
\left\langle%
q(i, \,t) \, q(j, \, t + \Delta) %
 \right\rangle %
 = \sum_{p=1}^N
\displaystyle\frac{2 \, \gamma(p) \, T(p)}{m^2(p)}
  \, \int_0^t \, d\tau \, W_{ip}(\tau) \,
 W_{jp}(\tau + \Delta)
\end{equation}
The generalization of expression (\ref{Fin_1}) for correlation functions taken
at different times is also obvious.

Further particular calculations will be done for the case, when the Langevin
forces acts only on extreme $1$-st and $N$-th particles and we'll present some
results for this case. We also consider for the definiteness that $\gamma(1) =
\gamma(N) = 1$. The solution of this problem we find as the solution of the
problem for eigenvalues (\ref{Vib_1}) and the system of linear equations
(\ref{Coeff_1}).

Note, that the standard matrix calculus with double precision (at the
diagonalization of matrix (\ref{Matrix_A}) give the relaxation times only of
the order $\tau \simeq 10^{14}$. Really, as the result we get the complex
value, and if the imaginary part (frequencies) $\simeq 1$, then it is
impossible to "catch" the real part $1/\tau < 10^{-14}$. This fact becomes
obvious if one returns back to the expression (\ref{Fin_1}). There stands a
value of the type $e^{i \, \omega \, t}$ in the  nominator, and if $\omega$ is
defined with the precision $\simeq 10^{-14}$, then it is not possible to use
times greater then $t > 10^{14}$. Nevertheless we can use $t = \infty$, because
of the damping the phase terms becomes to be zero.

Namely these long-living states correspond to highly localized states. It
follows from the well known fact, that there are localized states with eigen
frequencies $\sim 1$ and very small damping in long disordered chains.
Analogous localized states were discovered by P.Anderson \cite{And58} in the
diffusion problem (see also review \cite{Mot61} ).

Actually, in double summation (\ref{Fin_1}) the most dangerous is the case,
when in denominator $\lambda_{k_1} = \lambda^*_{k_2}$, what is the very small
number for the localized vibrational modes $\sim - 2 \, \tau_p$ .

But it turns out that the order of computations can be doubled. Namely, from
the equation ${\hat A} \, {\vec q} = \lambda \, {\vec q}$, rewritten in
components, it follows
\begin{equation}
\label{Eqa}%
m(p) \, x_k(p) \, \lambda^2_k = x_k(p-1) -2 \, x_k(p) + x_k(p+1) - \gamma(p) \,
\lambda_k \, x_k(p).
\end{equation}
From the equation (\ref{Eqa}) one can get (multiplying on complex conjugated,
summing and dividing real and imaginary parts) expression (valid only at
$\gamma(1) = \gamma(N) =1$):
\begin{equation}
\label{Eqb}%
\displaystyle\frac{1}{\tau_k} = - \displaystyle\frac{|x_k(1)|^2 + |x_k(N)|^2}{2
\, \sum_{i=1}^N \, m_i \, |x_i|^2}
\end{equation}
Thus there stands a value  of the order of unity in denominator. If eigenvector
is known with the precision of the order of $10^{-14}$, then the relaxation
times can be calculated with the precision $\sim 10^{-28}$, i.e. $\tau \sim
10^{28} $. It means that using the standard PC one can investigate the kinetics
of isotopically disordered chains for the times much greater then it is
available at any direct numerical simulation.

Thus, the limiting (by precision) stages of computations are the matrix
operations (\ref{Matrix_A}) and (\ref{second}).

Below we shall demonstrate some of the most essential results.

The spectrum of relaxation times depending on the chain length $N$ at $N \geq
100$ is shown in  Fig.~1. As we mainly interested in the long-time kinetics,
then the calculation of the relaxation times we've performed starting from
$\lg\tau > 7.5$ (up to this value the spectrum is irregular and it is
approximately the end point of relaxation times for the regular harmonic
chain). One can see, that the spectrum depends on $N$ for $N < 300$. Starting
from $N = 300$ the spectrum takes the constant value, as in the thermodynamical
limit this spectrum should be the constant value.

\begin{figure*}
\begin{center}
\includegraphics[height=0.27\textheight]{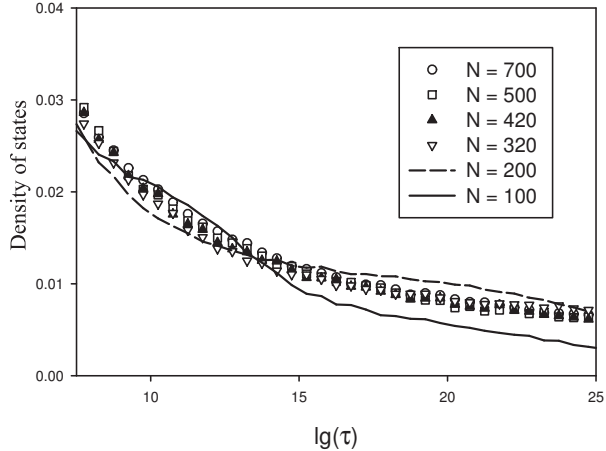}
\caption{\label{Fig1_e} The dependence of normalized distribution function for
relaxation times for different number of particles in the lattice. Solid curve
-- $N=100$, dashed line -- $N=200$. Points of different forms correspond to $N$
from $N =320$ to $N = 700$. }
\end{center}
\end{figure*}

The "tails" of relaxation times were approximated by several functional
dependencies $f(\lg(\tau))$, and the mean of logarithm of relaxation time in
all cases was calculated according to
\begin{equation}
\label{tau}
\left\langle%
\lg \tau
\right\rangle%
= \displaystyle\frac{\int_{7.5}^{\infty} \, \lg \tau \,f(\lg \tau) \, d\lg\tau
}{\int_{7.5}^{\infty} \, f(\lg \tau) \, d\lg\tau}
\end{equation}
It turned out that at this approach and the different choices of approximating
functions $\left\langle \lg\tau\right\rangle \approx 17$. Naturally, we
supposed that the relaxation law is valid for larger times, therefor the upper
limit in (\ref{tau}) is chosen to be the infinity.

For $N > 300$ the probability that $\lg \tau > 7,5$ is large then $1/3$. i.e.
area of the tail is large then $1/3$ from the total area under the curve of the
distribution function.

The largest time have vibrational modes, localized in the center of the chain.
It is connected with the fact that the corresponding vibrational wave functions
have negligibly small amplitudes at the Langevin sources (ends of the chain),
and these sources too long "drive" these modes. Naturally, that the maximum
relaxation times $\tau_{max}$ should grow with the growth of $N$. But the law
of growth of $\tau_{max}$ with $N$ is still unknown.

The temperature profiles along the chain for differen boundary temperatures are
shown in Fig.~2 for $N = 70 $, where our results are exact, for arbitrary
times. The profiles are shown for $\tau = 4, \, 6, \, 10,\, 14, \, \infty$ time
units.

As was shown above, because of extremely long life time of vibrational modes,
localized in the center of the lattice and having the relaxation times greater
than $>10^{28}$ for $N> 100$, the temperature in the center of the lattice is
calculated not very exactly. For the lattice of any size we can guarantee only
the results for $\sim 30-40$ particles from every end. It also means that the
numerical simulations (e.g., temperature distribution along the chain) for
large lattices are incorrect as the states localized in the center of the
lattice have no enough time to relax to the stationary state during the period
of time of the numerical simulation.

\begin{figure*}
\begin{center}
\includegraphics[height=0.27\textheight]{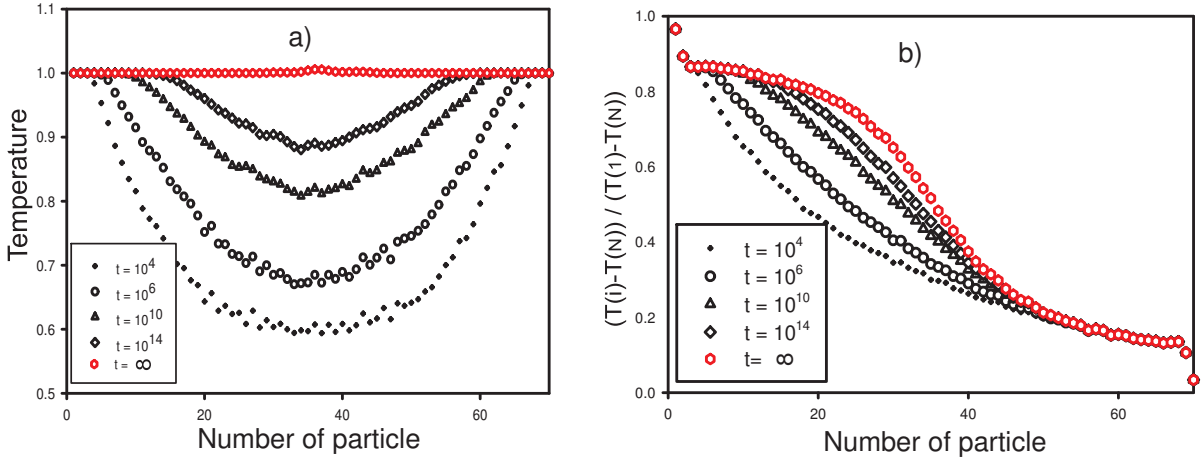}
\caption{\label{Fig2_e} The temperature profiles for the lattice of  70
particles at different boundary temperatures and times: a) $T(1) = T(N) = 1$;
b) $T(1) = 1, \,\,  T(N) = 0$}
\end{center}
\end{figure*}

Because of the thermal flow in the stationary state is a constant value along
the length of all the lattice, then it can be calculated in the vicinity of
ends, where, as was stated above, all our calculations are exact. It is obvious
that the system may not reach its stationary state. The specific coefficient of
thermal conductance $\kappa$ as dependent on the chain length is shown in
Fig.~3. $\kappa$ depends only on the temperature difference on the chain ends
and doesn't depend on the absolute temperature ($(T(1) + T(N))/2$), what is
obvious from (\ref{Corr_func_3}). Really, if the Langevin sources acts at
extreme atoms (1-st and $N$-th), then the expression (\ref{Corr_func_3}) can be
rewritten
\begin{equation}
\label{Corr_func_5}%
C_{ij} = \displaystyle\frac{2 \, T(1)}{m^2(1)} \, \int_0^t \, W_{i1} (\tau) \,
W_{j1} (\tau) \, d\tau + \displaystyle\frac{2 \, T(N)}{m^2(N)} \, \int_0^t \,
W_{iN} (\tau) \, W_{jN} (\tau) \, d\tau.
\end{equation}
Let's change the temperature at the ends in the same manner: $T(1) \rightarrow
T(1) +T, \quad T(N) \rightarrow T(N) +T $. Then the change in correlation
functions (\ref{Corr_func_5}) has the form of the correlator for equal
temperatures at the ends of the lattice:
\begin{equation}
\label{Corr_func_6}%
C_{ij} = \displaystyle\frac{2 \, T}{m_1^2} \, \int_0^t \, W_{i1} (\tau) \,
W_{j1} (\tau) \, d\tau + \displaystyle\frac{2 \, T}{m_N^2} \, \int_0^t \,
W_{iN} (\tau) \, W_{jN} (\tau) \, d\tau,
\end{equation}
and it is obvious, that any changes in the thermal flow and other additives,
depending on the temperature difference become zero.

As it is seen from Fig.~3, for not too long lattices, $\kappa$ grows as $\kappa
\simeq 2,4 \, \lg \, N$ upto $N \sim 300$,and then slowly decreases.
(Calculations for $N > 1000$ we couldn't perform because of computational
limitations).

\begin{figure*}
\begin{center} %
\includegraphics[height=0.27\textheight]{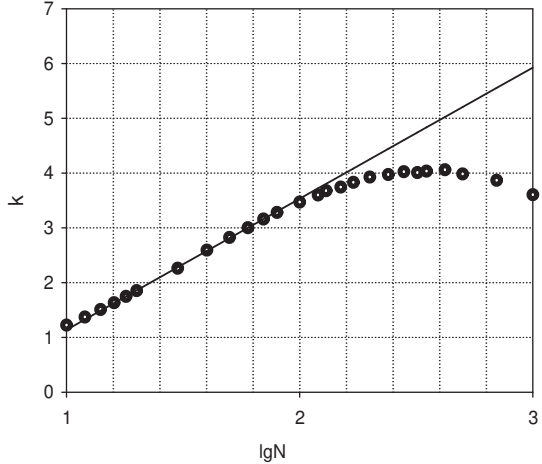}
\caption{\label{Fig3} The dependence of the coefficient of thermal conductance
on the chain length.  The linear fitting of the initial part of the curve
corresponds to $\kappa \simeq 2.4 \, \lg N$. }
\end{center}
\end{figure*}

Note, that all phonon states are localized for the 1D lattice \cite{Ber73},
\cite{Pas74}. It was proven for the electroconductance problem \cite{Ber73}
that the electroconductivity tends to zero as the absolute temperature also
tends to zero. As to our knowledge, there is no such a statement for the
thermal conductance of disordered 1D systems. But in the weak coupling
approximation, i.e. at $\gamma \ll 1$ it was supposed that for the rigid
boundary conditions $\kappa \sim 1/\sqrt{N}$, i.e. tends to zero \cite{Lep03a}.
(Recall the strange result that at free boundary conditions thermal conduction
diverges as $\kappa \sim \sqrt{N}$ \cite{Lep03a}).

It is probable that the slow decreasing of $\kappa$ for $N > 300$ obsereved in
Fig.~3 is the manifestation of the tendency of $\kappa$ to zero when $N \to
\infty$. We couldn't check this fact because of computational limitations.

The thermal conduction was calculated according to $\kappa = - J \, \Delta T$,
where $\Delta T$ was defined as the temperature difference on the lattice ends
(it is impossible to determine the temperature in the vicinity of a lattice
center, because of an existence of localized modes with extremely long time of
life). Therefor we restricted ourself by such a simple expression which is
valid by the order of magnitude.


In Conclusion we say that the solved problem partially continues the classical
works of Libowtz with co-workers \cite{Cas71}, \cite{Con74} on the dynamics of
1D disordered systems. We investigated the kinetic, dynamical and transport
properties of 1D isotopically disordered harmonic lattice. The accuracy of our
computations for all values, except the relaxation times, is limited by values
of the order of $10^{-14}$. We've succeeded to calculate the relaxation times
with the precision  $10^{-28}$. Our precision of solution is determined by
standard matrix operations performed on PC with double precision. Relaxation
times calculated by us many orderers of magnitude exceed any value obtained in
direct numerical simulations. We've got non-monotonic dependence of thermal
conduction which is dependent on $N$ with maximum at $N \sim 300$. The
stationary state is unique but times of its achievement grow with the increase
in chain length. Because of additivity of all results on temperature
(\ref{Fin_1}), the detailed calculations can be done at single chosen
temperature at the chain ends.

\vspace{0.3 cm}

This work was supported by the RFBR (02-02-16205, 04-03-32119 , 04-02-17306)
and the Program of Presidium of Russian Academy of Sciences "Fundamental
problems of physics and chemistry of nanosized systems and nanomaterials".


\begin{references}

\baselineskip 15 pt

\bibitem{Lep03a} S.Lepri, R.Livi, and A.Politi, Phys.Reports, {\bf 377}, 1 (2003).

\bibitem{Li_01} B.Li, H.Zhao,and B.Hu, Phys.Rev.Let., {\bf 86}, 63 (2001).

\bibitem{Lep98} S.Lepri, R.Livi, and A.Politi, Europhys.Lett., {\bf 43}, 271 (1998).

\bibitem{Lep97} S.Lepri, R.Livi, and A.Politi, Phys.Rev.Lett.,{\bf78}, 1896 (1997).

\bibitem{Sav02} A.V.Savin, G.P.Tsironis, and A.V.Zolotaryuk, Phys.Rev.Lett.
{\bf 88}, 154301 (2002).

\bibitem{Hat99} T.Hatano, Phys.Rev.E, {\bf 59}, R1 (1999).

\bibitem{Dha01a} A.Dhar, Phys.Rev.Lett. {\bf 86}, 3554 (2001).

\bibitem{Ern91}   M.H.Ernst, Physica D, {\bf47}, 198 (1991).

\bibitem{Nar02} O.Narayan, S.Ramaswamy, Phys.Rev.Lett., {\bf 89}, 200601
(2002).



\bibitem{Gar01} P.L.Garrido, P.I.Hurtado, and B.Nadrowski,
Phys.Rev.Lett., {\bf 86}, 5486 (2001).

\bibitem{Gia00} C.Giardina, R.Livi, A.Politi, and M.Vassallo,
Phys.Rev.Lett., {\bf 84}, 2144 (2000).

\bibitem{Gen00} O.V.Gendelman, A.V.Savin, Phys.Rev.Lett. {\bf 84}, 2381 (2000).

\bibitem{Yan03} L.Yang, P.Grassberg, arXiv:cond-mat/0306173.

\bibitem{Dha01b} A.Dhar, Phys.Rev.Lett. {\bf 86}, 5882 (2001).

\bibitem{Pro00} T.Prosen, D.K.Campbell, Phys.Rev.Lett., {\bf 84}, 2857
(2000).

\bibitem{Mar00} S.Maruyama, in {\it Advances in Numerical Heat
Transfer}, vol.2, 189 (2000) W.J.Minkowjcz and E.M.Sparrow (Eds.), Taylor \&
Francis, NY, 2000.

\bibitem{Mar02} S.Maruyama, Physica B, {\bf 323}, 193 (2002).

\bibitem{Mar03} S.Maruyama, Microscale Thermophysical
Engineering, {\bf 7}, 41 (2003).

\bibitem{Cao04} J.X.Cao, X.H.Yan, Y.Xiao, and J.W.Ding, Phys.Rev.B, {\bf 69},
073407 (2004).

\bibitem{Dys53} F.J.Dyson, Phys.Rev. {\bf 92}, 1331 (1953).

\bibitem{For03} P.J.Forrester, N.C.Snaith,, and J.J.M.Verbaarschot,
arXiv:cond-mat/0303207.

\bibitem{And58} P.W.Anderson, Phys.Rev. {\bf 109}, 1492 (1958).

\bibitem{Mot61} N.Mott, W.Twose,  Adv.Phys., {\bf 10}, 107 (1961), .

\bibitem{Ber73} V.L.Berezinskii, JETP {\bf 65}, 1251 (1973) (in Russian).

\bibitem{Pas74} L.A.Pastur, Z.P.Fel'dman, JETP {\bf 67}, 487 (1974) (in Russian).

\bibitem{Cas71} A.Casher, J.L.Lebowitz, J.Math.Phys. {\bf 12}, 1701 (1971).

\bibitem{Con74} A.J.O'Connor, J.L.Lebowitz, J.Math.Phys. {\bf 15}, 692
(1974).

\end{references}
\end{document}